\documentclass[a4paper,11pt,,dvipsnames,table]{article}
\usepackage{pos}
\usepackage{makecell}

\definecolor{light-gray}{HTML}{E5E4E2}
\definecolor{light-cyan}{HTML}{E0FFFF}

\title{Completing Aganta Kairos: Capturing Metaphysical Time on the Seventh Continent}
 \ShortTitle{Aganta Kairos}
 \author{The IceCube Collaboration \\{\normalsize \normalfont(a complete list of authors can be found at the end of the proceedings)}}

\emailAdd{jim.madsen@icecube.wisc.edu}
\emailAdd{agantakairos@icloud.com}
\emailAdd{christian.spiering@desy.de}

\abstract{We present an overview of the art project Aganta Kairos (To Fish the Metaphysical Time). This project celebrates the neutrino—the ghost particle—which scientists consider a cosmic messenger and the artist regards as a link between people who care about their relationship to the cosmos and question their origins. The artwork is based on a performance of celebration and seeks to build a human community that encompasses different knowledge domains and interpretations of the universe. This intersection of knowledge is realized during the performance of placing a plaque, held with witnesses, and during subsequent exhibitions. Images, sounds, videos, and sculpture testify to the diversity of approaches to questioning our origins, ranging from traditional western science to ancient shamanism. The sites were selected for their global coverage and, for the South Pole, Mediterranean, and Lake Baikal, their connection to ongoing neutrino experiments. In December 2020, a plaque was installed at the South Pole IceCube Laboratory, the seventh and final site. We provide examples of images and links to additional images and videos.

\vspace{4mm}

{\bfseries Corresponding authors:}
Jim Madsen$^{1*}$, Laurent Mulot$^{2}$, Christian Spiering$^{3}$\\
{$^{1}$ \itshape University of Wisconsin-Madison, Wisconsin, USA}\\
{$^{2}$ \itshape Mulot Studio, Lyon, France}\\
{$^{3}$ \itshape DESY, Zeuthen  Germany\\[4mm]}
$^*$ Presenter

\FullConference{37$^{\rm{th}}$ International Cosmic Ray Conference (ICRC 2021)\\
		July 12th -- 23rd, 2021\\
		Online -- Berlin, Germany}
}		

\begin{document}

\maketitle

\section{Introduction}
\begin{figure}[h!]
  \begin{center}
  \includegraphics[width=4in]{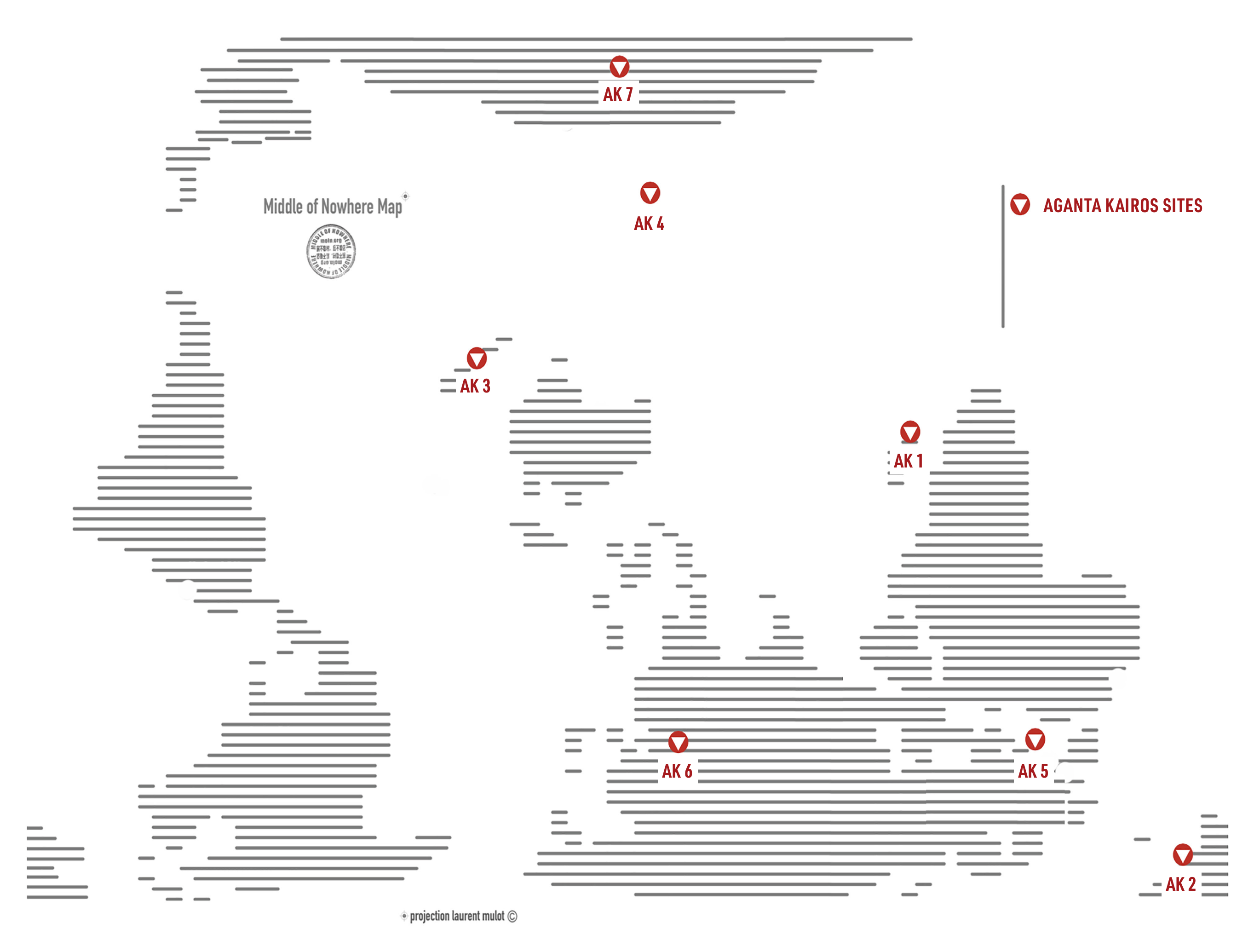}
  \caption{The seven locations (red symbols) where plaques have been installed for the Aganta Kairos project with the South Pole at the top of the map.}
  \label{fig:map}
   \end{center} 
\end{figure}

In 2012, artist \href{mofn.ens-lyon.fr/}{Laurent Mulot}~\cite{mulot} started work on his \href{mofn.ens-lyon.fr/aganta-kairos-us.html}{Aganta Kairos (To Fish the Metaphysical Time)} project~\cite{agantakairos}. In the first two years, the passage of neutrinos, phantom particles considered essential cosmic messengers by astrophysicists, was celebrated on the coasts of five oceans. 
Most neutrinos leave no trace, but with great effort, partners to this work, such as the ANTARES and KM3NeT/ORCA underwater telescopes  
in the abyss of the Mediterranean Sea off the island of Porquerolles, are able to capture a few. In 2018, the artist continued his work on the shores of Lake Baikal, the ``6th ocean,'' with the scientists in charge of the Baikal Neutrino Telescope.  The ceremony at the final site, the IceCube Neutrino Observatory at the South Pole, took place at the end of 2020. The map in Figure \ref{fig:map} shows the location of the sites with summary information in Table~\ref{tab:sites}. 

Aganta Kairos began with conversations with members of the ANTARES experiment in France.  The goal for the artist was to use the ghostly, cosmic neutrino to weave together people around the planet who have disparate ways of knowing. A sky map of the neutrinos detected deep under the water by ANTARES from 2008 to 2011 inspired Mulot to focus on the shores of the six oceans.  Lake Baikal, also a site of a long running neutrino project, was included in the group as the largest volume and deepest fresh water lake in the world, whose formation was similar to that of the oceans. The sites were selected to mix both places where neutrinos had passed through the Earth according to the ANTARES sky map and the scientific facilities where detection takes place, and to capture different magical, cultural and scientific traditions. Mulot visited all sites except the South Pole.  For example, in Madagascar he met with inhabitants of the village of Efoetsy and told the story of the passage and detection of a neutrino, and proposed a celebration. The cosmogony of this Malagasy tribe (Mahafaly) is very rich and maintains a particularly interesting dialogue with the celestial vault that was incorporated into the Aganta  Kairos exhibition.  The witnesses at each site do not use the same approach nor do they aim at the same objectives, but they all investigate our relationship to the cosmos and the universe that encompasses us. The process was repeated in New Zealand with the Maoris, and in Greenland with the Inuits. At Lake Baikal, the meeting with the scientists was enriched by discussions with a shaman; Siberia is the origin of shamanism. 

\begin{figure*}[ht!]
\begin{center}
\includegraphics[width=5in]{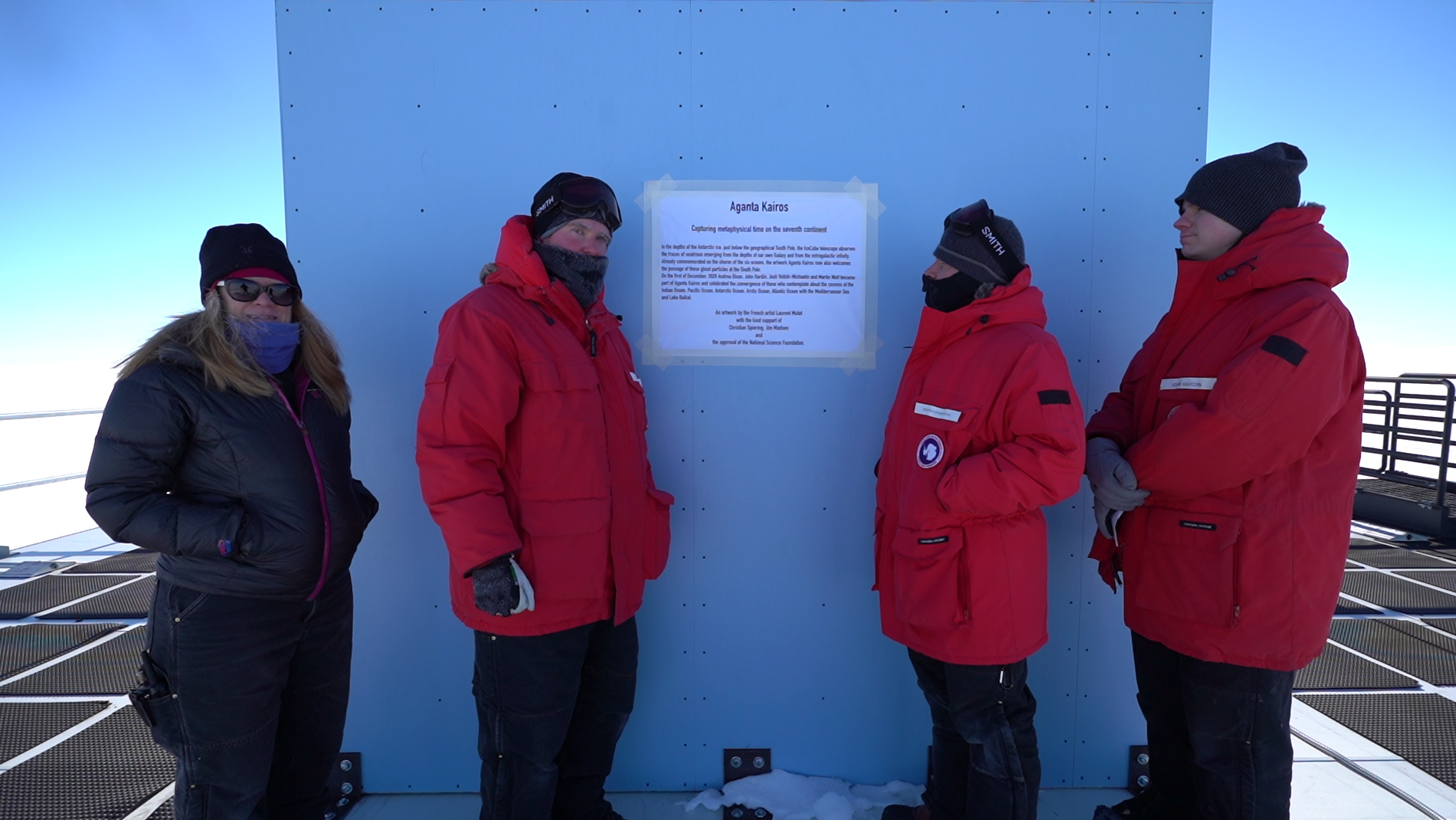}
\includegraphics[width=2.5in]{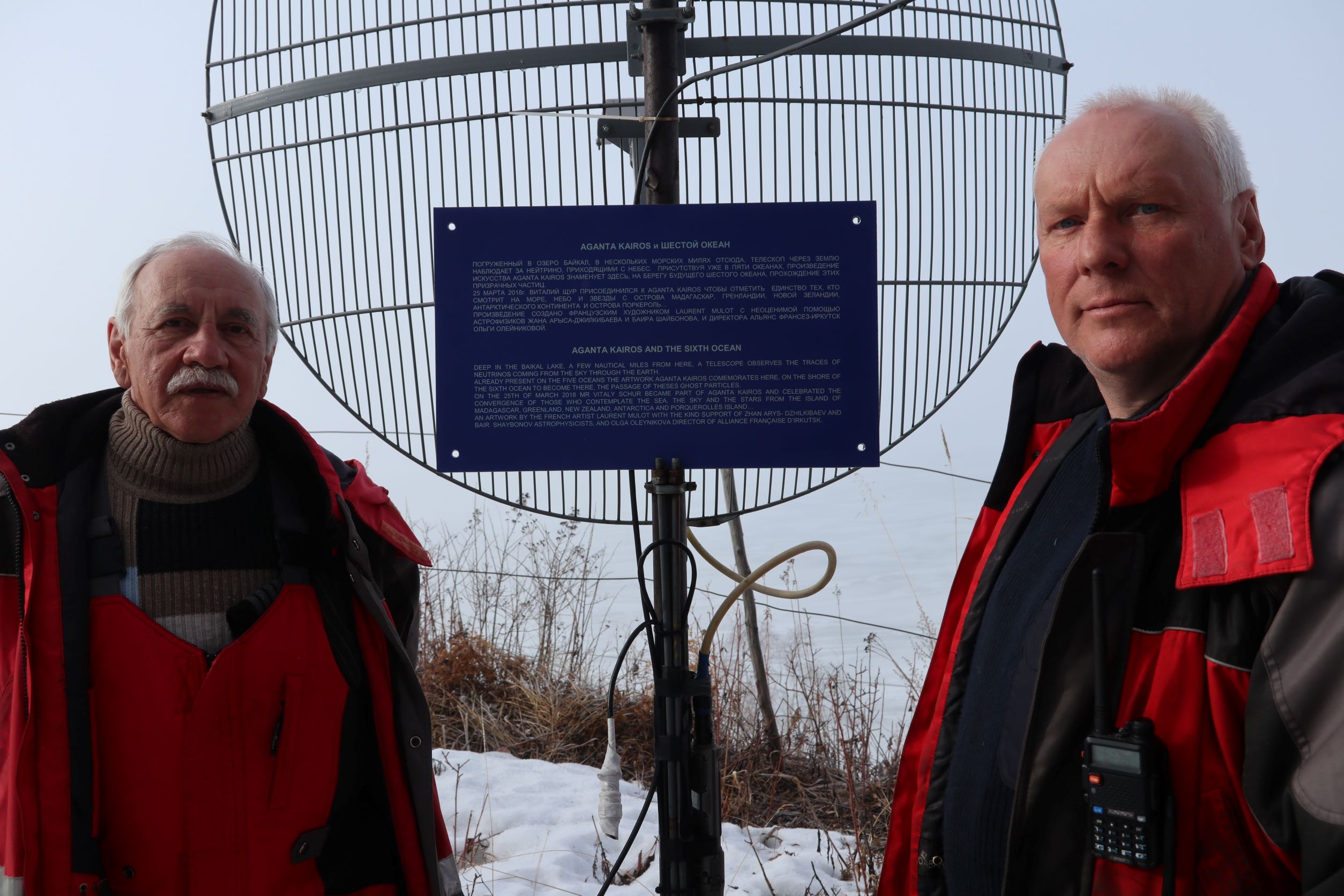}
\includegraphics[width=2.5in]{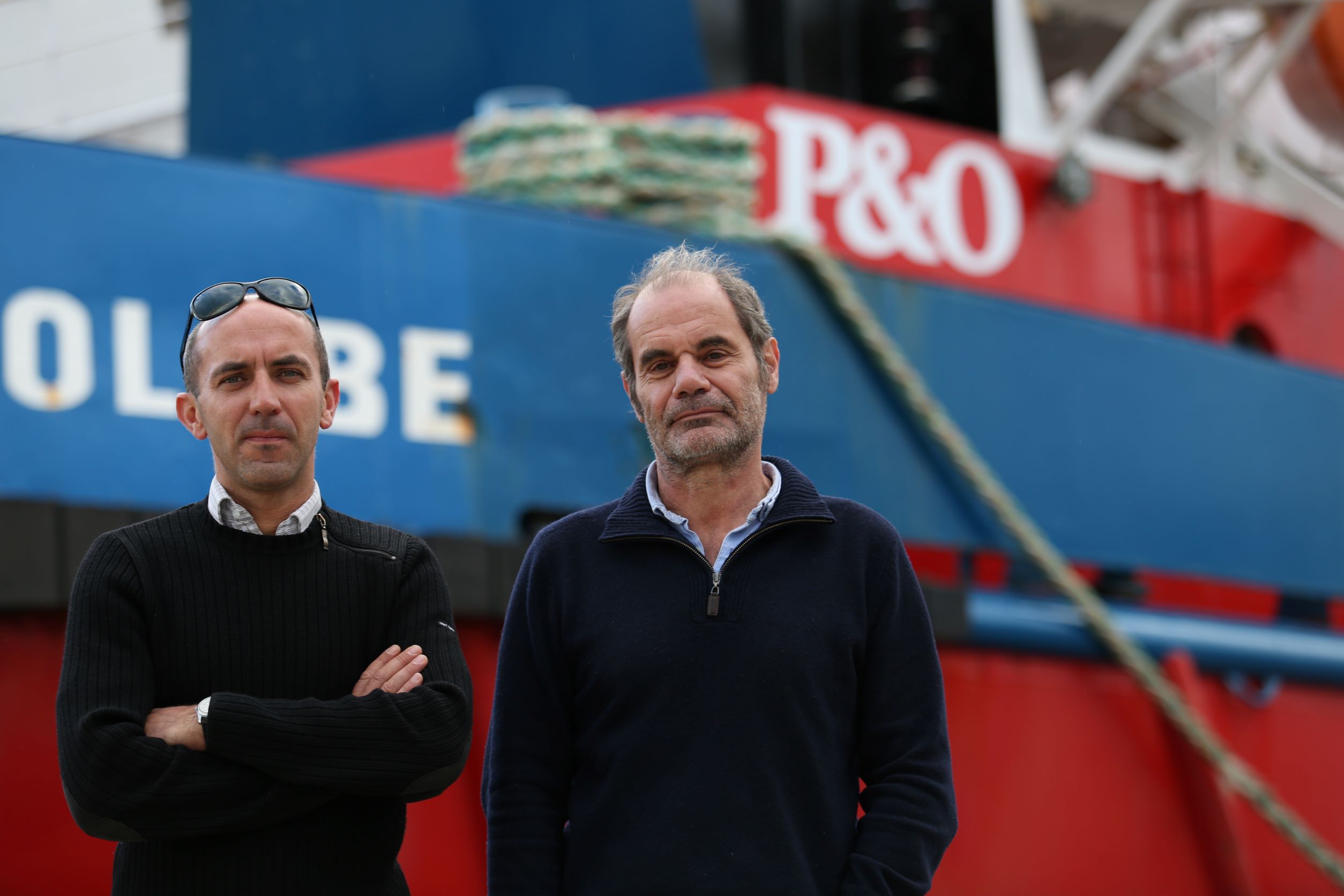}
\includegraphics[width=2.5in]{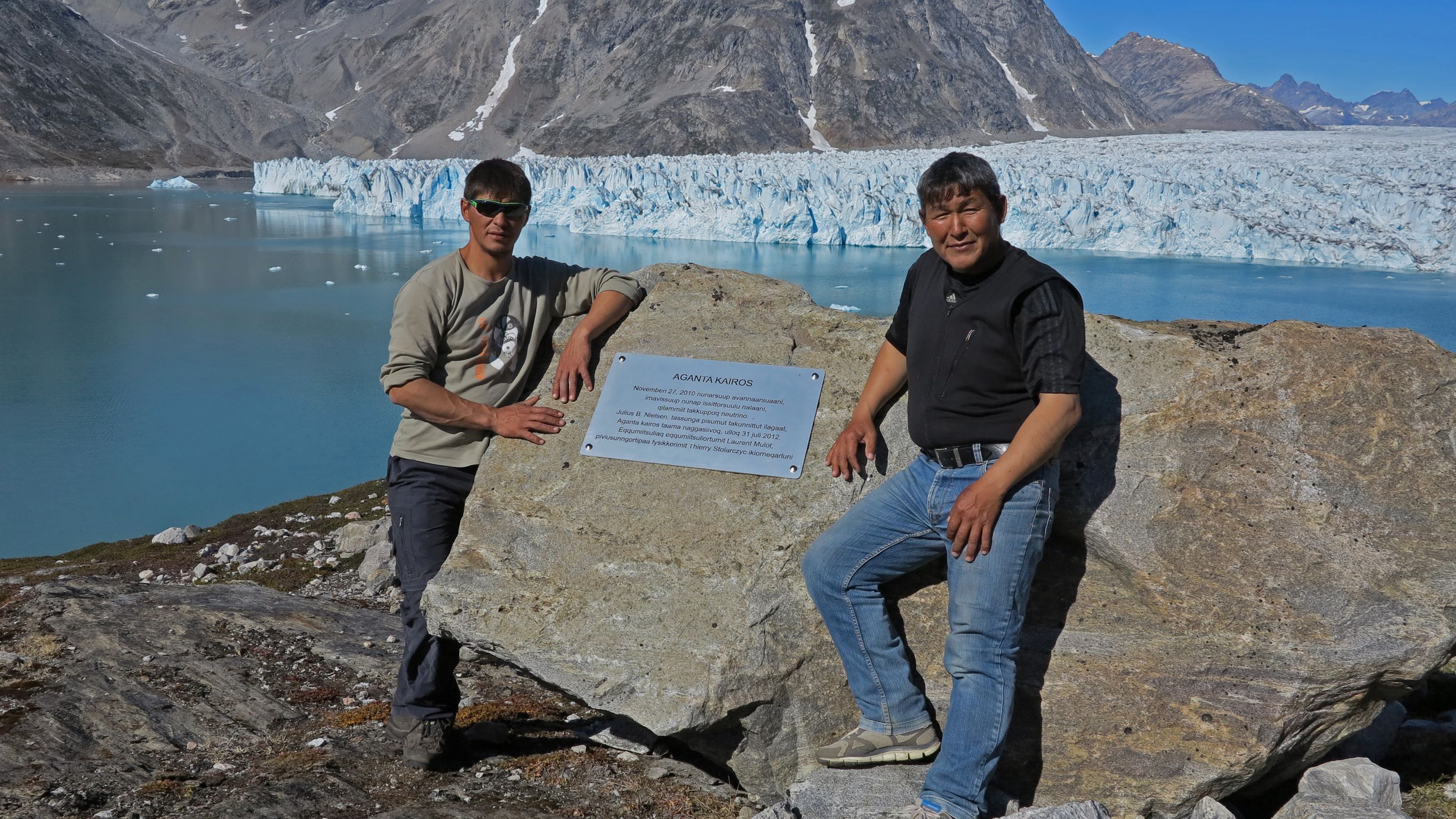}
\includegraphics[width=2.5in]{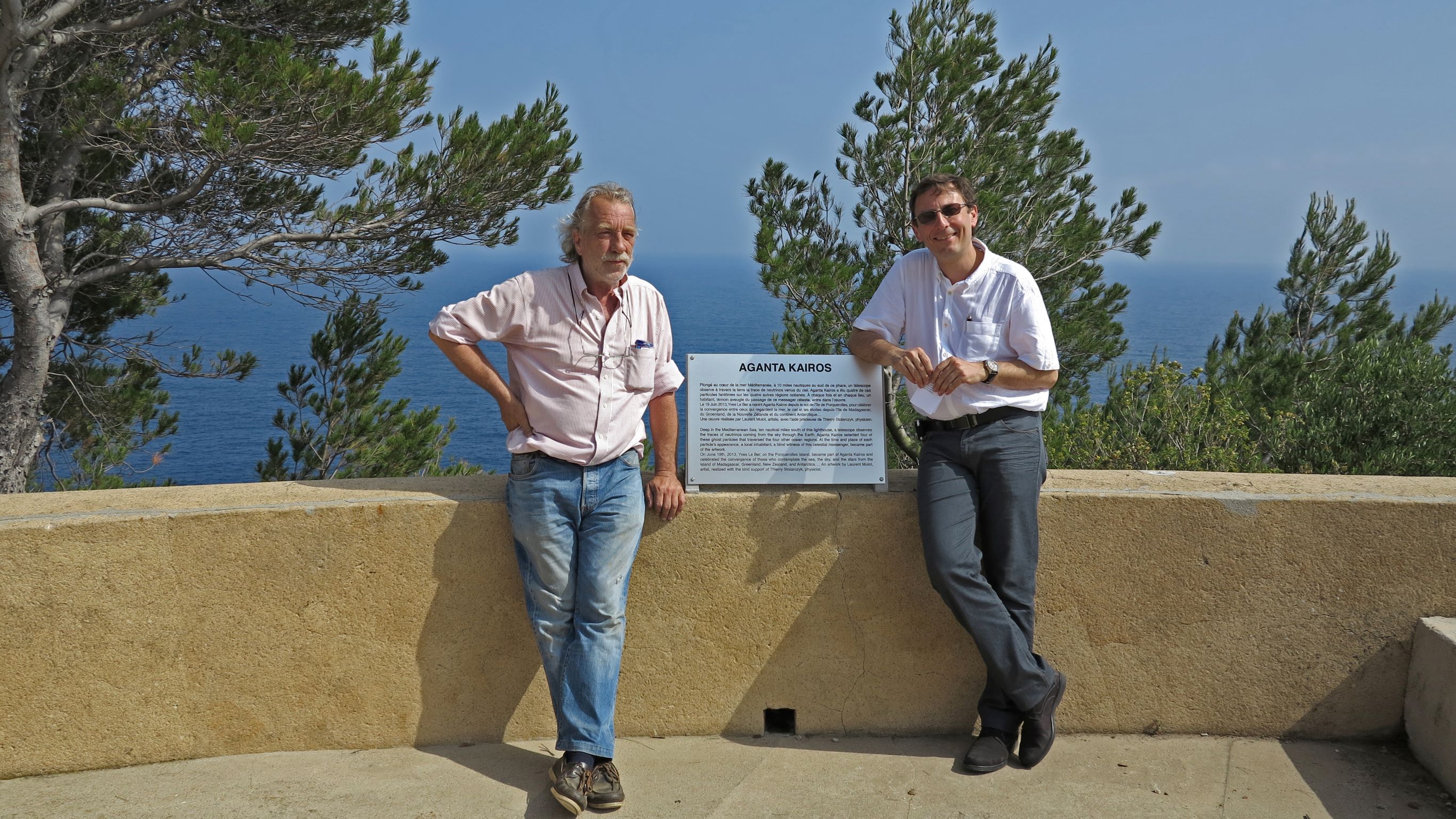}
\includegraphics[width=2.5in]{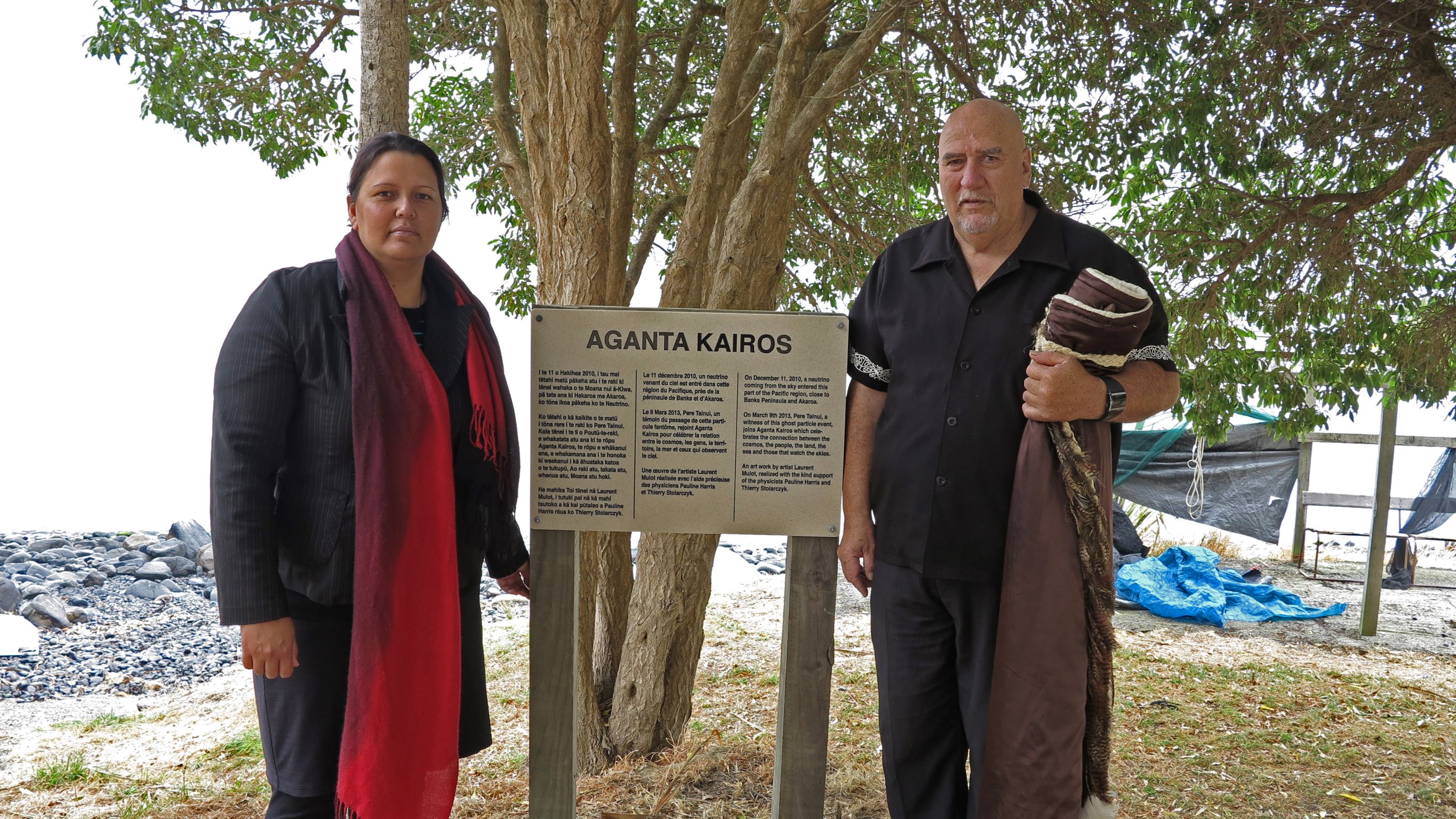}
\includegraphics[width=2.5in]{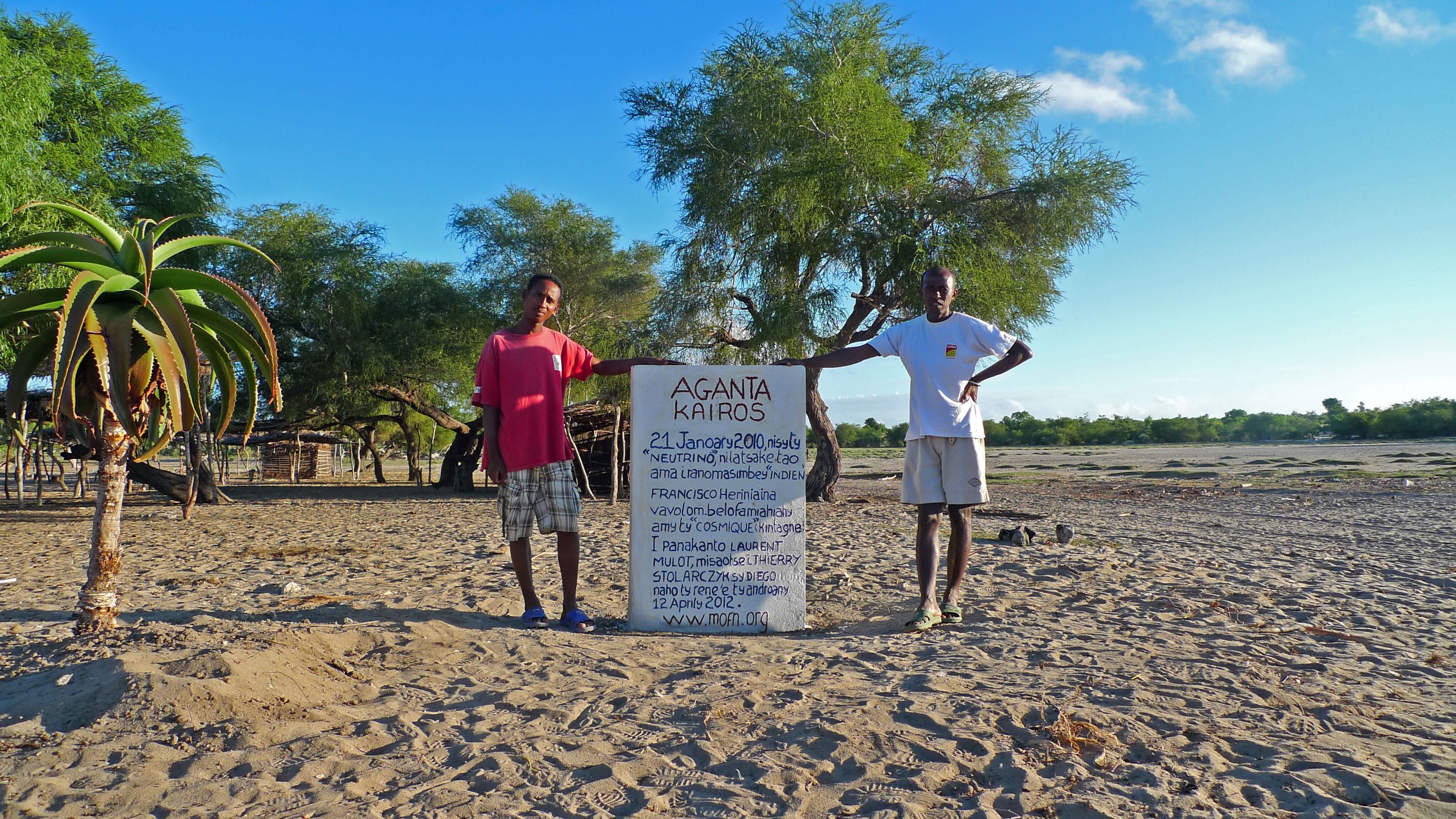}
\caption{The witnesses in reverse chronological order of the installation date starting from the top, then from left to right in each row. See Table~\ref{tab:sites} for information about each site. There is no plaque for the Austral Ocean site; the ceremony was performed on a ship en route to Antarctica. At the coordinates indicated in Table~\ref{tab:sites}, water was drawn from the Austral (also known as the Southern or Antarctic) Ocean.}
\label{fig:sites}
\end{center}
\end{figure*}

The resulting work, Aganta Kairos~\cite{agantakairos}, explores questions about our cosmic origins and recognizes different approaches to addressing them, ranging from science to mythology to the magical spirit of shamanism. At each site, Mulot realizes a performance with local witnesses 
of the invisible, including the installation of a commemorative plaque, as shown in Figure~\ref{fig:sites}. Seeing in a new way, whether from an advance in technology or by utilizing a different cultural framework, enables discoveries, revealing what was previously hidden. Using the antisocial neutrino that seldom interacts to connect cultures and explore different ways of knowing is an ironic artistic twist.  A particle that is almost nothing at all is being used by the artist Mulot to unite people and places around the world in the eternal search for deeper understanding. With 7 locations, including three of the largest neutrino detection experiments, Aganta Kairos is finished. Here we describe how IceCube got involved and the subsequent effort to complete the installation of the plaque at the South Pole.  

\begin{table}
    \centering
   \footnotesize
    \begin{tabular}{ m{4.5cm}|m{1.8cm}|m{2.5cm}|m{2.3cm}|m{1.8cm}}
        \hline
        \textbf{ Witnesses} & \textbf{Lat/ Long} &  \textbf{Date} &  \textbf{Description} & \textbf{ Location} \\
        \hline
        \hline
         Andrea Dixon, Martin Wolf \newline Josh Veitch-Michaelis, John Hardin & 90° S \newline    0° E & December 21, 2020 & IceCube Neutrino \newline Observatory & South Pole, \newline  Antarctica\\
        \hline
         Zhan-Ayrs Dzhilkibaev \newline Vitaly Schur & 51°48'54" N \newline 104°50'07" E & March 25. 2018 &  Baikal Neutrino \newline Telescope & Lake Baikal, \newline Russia\\
        \hline
        Stanislas Devorsine \newline Benoit Hebert &  63°28' S \newline 139°48' E & February 7, 2014 & & Austral Ocean\\
        \hline
         Julius Nielson \newline Gerd lgnatussen & 66°23'24" N \newline 36°47'46" W & July 31, 2012 & Sermilik Fjord & Greenland\\
        \hline
        Yves Le Ber \newline Thierry Stolarczyk & 42°59'00" N \newline 6°12'24" E & June 19, 2013 & Antares Neutrino \newline Telescope & Porquerolles Island, France\\
        \hline
         Pauline Harris \newline Pere Tanui & 43°50'07" S \newline  172°56'44" E&March 9. 2013 & Sacred place of \newline Onuku & Marae, \newline New Zealand\\
        \hline
        Francisco Heriniania \newline Mr. President & 24°04'41" S \newline 43°41'57" E & April 12, 2012  & & Efoetsy, Madagascar\\
        \hline
    \end{tabular}
    \caption{A listing of the names of the witnesses, geographic coordinates, installation dates, descriptions, and locations for the seven sites in reverse chronological order.  Mr. "President" is the name the chief of the village of Efoetsy goes by.}
    \label{tab:sites}
\end{table}

\section{The South Pole Connection}
 Aganta Kairos grew organically by identifying potential partners who were connected to and cared about their origins and then working with those willing to collaborate.   The key action is the inauguration of the plaque celebrating the neutrino, the ghost particle, which is considered by scientists essential to understanding the universe and taken by the artist as a link between people. IceCube not only added the third and largest neutrino observatory to the project, but also provided geographic balance through its location at the South Pole.  

While neutrinos reach the South Pole from all directions with ease, scientist and support staff have only one route, via the US Antarctic Program run by the US National Science Foundation (NSF).  Even prior to COVID, only a few hundred people per year made it all the way to bottom of the Earth, and those that did adhered to guidelines that limit what can be transported to South Pole.  To ensure that there would be no problems placing the plaque and promoting IceCube's inclusion,  permission to participate in the art project was sought and received from the NSF.

\section{Preparation and Installation}
In a typical season, IceCube sends about a dozen people to the South Pole.  
Coordinating their activities is an exercise in resilience; thorough planning is needed but the ability to go with the flow is equally important. For the South Pole, travel takes place during the austral summer, from mid-November to mid-February or so. Cargo delays combined with weather or mechanical issues that change travel dates ensure that the best-laid plans should be treated as guides rather than absolutes---reasonable ways to accomplish what is needed but unlikely to be executed exactly as written. Travel restrictions due to a worldwide pandemic were not in the plan and required NSF to rethink who and what was essential for the Antarctic program in the 2020-2021 season.   

In a more normal season, getting a bronze plaque roughly the size of a sheet of A3 paper (0.30mx0.42m) to the South Pole would not be difficult.  If it was available early enough, it could be included in the IceCube cargo for the season. If it was not ready, the other option would be to have someone who was deploying carry it with them by hand in their luggage. For this project, neither option was viable to transport the plaque because all cargo and deployments had been canceled except those deemed essential for continuing operation of the station and select science projects.  For IceCube, that meant deploying only the two replacement winterovers, who are responsible for keeping the detector running, and a very minimal amount of critical cargo.  Since the larger C-130 planes would not be flying to the South Pole, carry-on weight allowances were reduced, leaving no options to transport a bronze plaque. 

\begin{figure}[h]
  \begin{center}
  \includegraphics[width=4.75in]{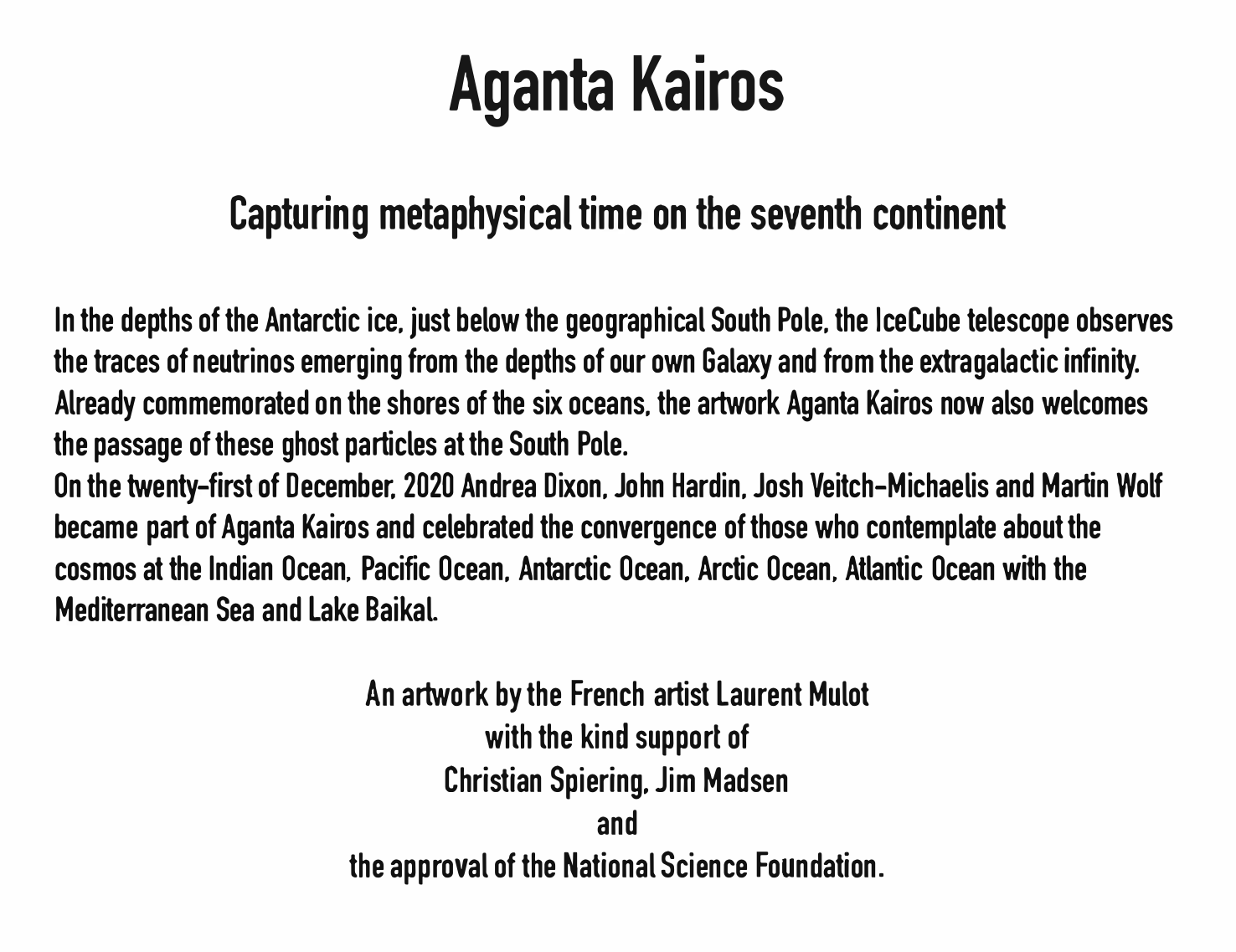}
  \caption{The text for the plaque commemorating IceCube becoming a part of Aganta Kairos.}
  \label{fig:plaque}
   \end{center} 
\end{figure}

\begin{figure*}[ht!]
\begin{center}

\includegraphics[width=5.45in]{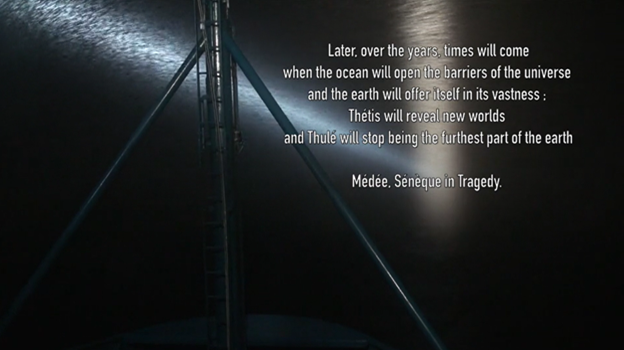}
\includegraphics[width=2.7in]{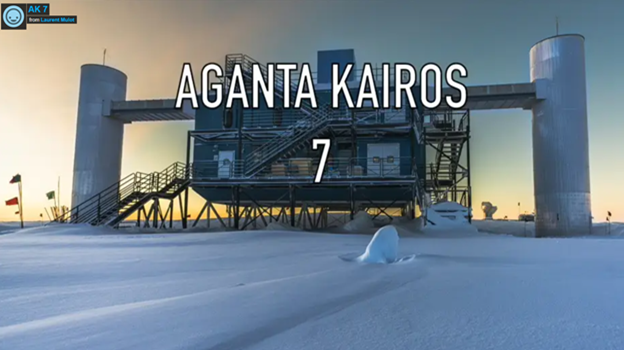}
\includegraphics[width=2.7in]{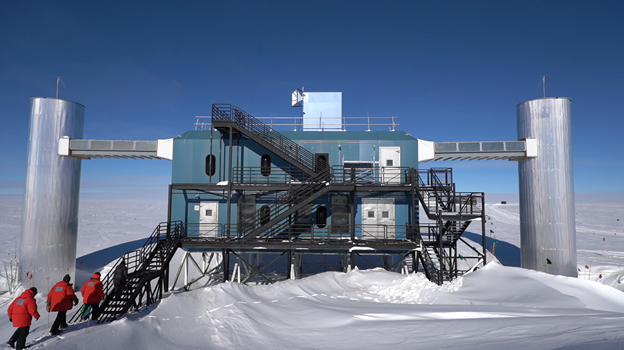}
\includegraphics[width=2.7in]{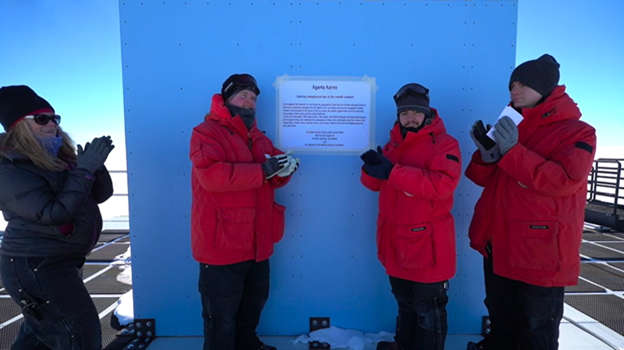}
\includegraphics[width=2.7in]{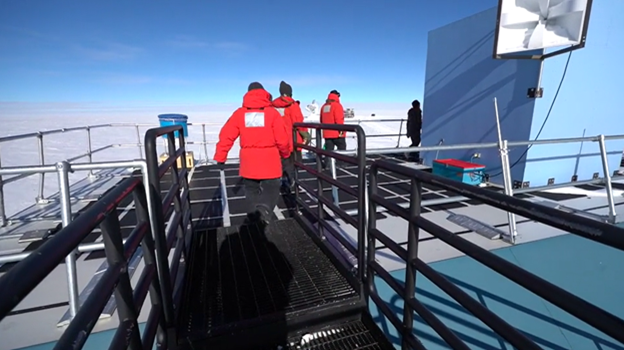}

\caption{Outtakes from the video~\cite{installation} documenting the installation ceremony at the IceCube Laboratory (ICL) at the South Pole. The top image shows text from the Roman tragedy Medea by Seneca. Starting from the middle left and going clockwise, the opening shot shows the ICL. The next two pictures show the witnesses climbing the stairs and then walking on the roof of the ICL.  The lower left pictures shows the witnesses applauding after the installation was complete.   The paper plaque will be replaced with permanent bronze version when possible.  }
\label{fig:outtakes}
\end{center}
\end{figure*}

We produced an alternative plan to deal with these restrictions. A temporary, lightweight cloth version of the plaque was made that could easily fit in a suitcase. But getting the cloth plaque to the IceCube winterovers turned out to be another challenge because of quarantine requirements. In the end, the cloth plaque was packed in a suitcase as planned; however, that luggage did not make it to the South Pole until one of the last flights of the season, well after the December 2020 installation ceremony.  Undeterred, our witnesses, IceCube winterovers John Hardin, Josh Veitch-Michaelis and Martin Wolf, and South Pole station manager Andrea Dixon and their volunteer videographer, South Pole Telescope winterover Geoffrey Chen, found a way to succeed.  They printed an even more temporary plaque on paper as you can see in the photo in Figure~\ref{fig:sites}.  The text for the plaque is shown in Figure~\ref{fig:plaque}.  A few outtakes from the video~\cite{installation} commemorating IceCube joining Aganta Kairos
filmed on December 21, 2020, when the temperature was -26\textdegree~C (-15\textdegree~F) are shown in Figure~\ref{fig:outtakes}.

\section{Exhibitions and Catalog}
Aganta Kairos has already appeared in several exhibitions while it has been under development, including the International Biennale of Lyon in 2013/14, and will be presented during the next Biennale at Bleu du Ciel Gallery Lyon in its final version in the fall of 2022. 
For this exhibition, a monumental video work showing seven synchronous videos of the seven Aganta Kairos inaugurations will be presented as well as a sculpture mixing water from the seven sites. 

A catalog entitled \textbf{Aganta Kairos: from the six oceans to the seventh continent} 
will be published in 2022.  It will include sections written by Mulot's scientific partners, text by the contemporary art critic Paul Ardenne, 
and selections from the series of photographs of Aganta Kairos. 

\section{Summary and Outlook}
Science provides one way to explore the universe, but it is not the only path to its understanding. Just as multimessenger astronomy and astrophysics bring together different ways to view an object, enabling greater insight, other disciplines and cultures can contribute to answering questions about what it means to be human and to ponder our place in the universe. Working with passionate people is always enlightening, more so when their approach to viewing and understanding uses unfamiliar methods or ideas. The IceCube Collaboration has had the opportunity to work with a number of artists, and each has seen something new in what we do and communicated it in a way that reaches new audiences. Working with artist Mulot has been particularly rewarding, as his work, Aganta Kairos, unites those searching for answers about the universe using messages from neutrinos with those using other ways of knowing.  We look forward to the opportunity to see this work exhibited and to be able to react and reflect in real time with others who ponder how to detect and decipher messages from the hidden universe.  

\section{Key Contributions}
We thank Dr.~Vladimir Papitashvili (Program Director, Astrophysics \& Geospace Sciences, NSF) and Elaine Hood (Communications, Antarctic Support Contract) for their help in realizing this project.
We thank Andrea Dixon (US Antarctic Program), John Hardin (WIPAC), Josh Veitch-Michaelis (WIPAC), Martin Wolf (WIPAC) and Geoffrey Chen (University of Chicago) for performing the inauguration ceremony and installing the plaque.

\clearpage
\section*{Full Author List: IceCube Collaboration}




\scriptsize
\noindent
R. Abbasi$^{17}$,
M. Ackermann$^{59}$,
J. Adams$^{18}$,
J. A. Aguilar$^{12}$,
M. Ahlers$^{22}$,
M. Ahrens$^{50}$,
C. Alispach$^{28}$,
A. A. Alves Jr.$^{31}$,
N. M. Amin$^{42}$,
R. An$^{14}$
% ...
\\

\noindent
% ...
$^{12}$ Universit{\'e} Libre de Bruxelles, Science Faculty CP230, B-1050 Brussels, Belgium \\
% ...
$^{14}$ Department of Physics and Laboratory for Particle Physics and Cosmology, Harvard University, Cambridge, MA 02138, USA \\
% ...
$^{17}$ Department of Physics, Loyola University Chicago, Chicago, IL 60660, USA \\
$^{18}$ Dept. of Physics and Astronomy, University of Canterbury, Private Bag 4800, Christchurch, New Zealand \\
% ...
$^{22}$ Niels Bohr Institute, University of Copenhagen, DK-2100 Copenhagen, Denmark \\
% ...
$^{28}$ D{\'e}partement de physique nucl{\'e}aire et corpusculaire, Universit{\'e} de Gen{\`e}ve, CH-1211 Gen{\`e}ve, Switzerland \\
% ...
$^{31}$ Karlsruhe Institute of Technology, Institute for Astroparticle Physics, D-76021 Karlsruhe, Germany  \\
% ...
$^{42}$ Bartol Research Institute and Dept. of Physics and Astronomy, University of Delaware, Newark, DE 19716, USA \\
% ...
$^{50}$ Oskar Klein Centre and Dept. of Physics, Stockholm University, SE-10691 Stockholm, Sweden \\
% ...
$^{59}$ DESY, D-15738 Zeuthen, Germany \\
% ...

\subsection*{Acknowledgements}

\noindent
USA {\textendash} U.S. National Science Foundation-Office of Polar Programs,
U.S. National Science Foundation-Physics Division,
U.S. National Science Foundation-EPSCoR,
Wisconsin Alumni Research Foundation,
Center for High Throughput Computing (CHTC) at the University of Wisconsin{\textendash}Madison,
Open Science Grid (OSG),
Extreme Science and Engineering Discovery Environment (XSEDE),
Frontera computing project at the Texas Advanced Computing Center,
U.S. Department of Energy-National Energy Research Scientific Computing Center,
Particle astrophysics research computing center at the University of Maryland,
Institute for Cyber-Enabled Research at Michigan State University,
and Astroparticle physics computational facility at Marquette University;
Belgium {\textendash} Funds for Scientific Research (FRS-FNRS and FWO),
FWO Odysseus and Big Science programmes,
and Belgian Federal Science Policy Office (Belspo);
Germany {\textendash} Bundesministerium f{\"u}r Bildung und Forschung (BMBF),
Deutsche Forschungsgemeinschaft (DFG),
Helmholtz Alliance for Astroparticle Physics (HAP),
Initiative and Networking Fund of the Helmholtz Association,
Deutsches Elektronen Synchrotron (DESY),
and High Performance Computing cluster of the RWTH Aachen;
Sweden {\textendash} Swedish Research Council,
Swedish Polar Research Secretariat,
Swedish National Infrastructure for Computing (SNIC),
and Knut and Alice Wallenberg Foundation;
Australia {\textendash} Australian Research Council;
Canada {\textendash} Natural Sciences and Engineering Research Council of Canada,
Calcul Qu{\'e}bec, Compute Ontario, Canada Foundation for Innovation, WestGrid, and Compute Canada;
Denmark {\textendash} Villum Fonden and Carlsberg Foundation;
New Zealand {\textendash} Marsden Fund;
Japan {\textendash} Japan Society for Promotion of Science (JSPS)
and Institute for Global Prominent Research (IGPR) of Chiba University;
Korea {\textendash} National Research Foundation of Korea (NRF);
Switzerland {\textendash} Swiss National Science Foundation (SNSF);
United Kingdom {\textendash} Department of Physics, University of Oxford.\scriptsize
\noindent
R. Abbasi$^{17}$,
M. Ackermann$^{59}$,
J. Adams$^{18}$,
J. A. Aguilar$^{12}$,
M. Ahlers$^{22}$,
M. Ahrens$^{50}$,
C. Alispach$^{28}$,
A. A. Alves Jr.$^{31}$,
N. M. Amin$^{42}$,
R. An$^{14}$
% ...
\\

\noindent
% ...
$^{12}$ Universit{\'e} Libre de Bruxelles, Science Faculty CP230, B-1050 Brussels, Belgium \\
% ...
$^{14}$ Department of Physics and Laboratory for Particle Physics and Cosmology, Harvard University, Cambridge, MA 02138, USA \\
% ...
$^{17}$ Department of Physics, Loyola University Chicago, Chicago, IL 60660, USA \\
$^{18}$ Dept. of Physics and Astronomy, University of Canterbury, Private Bag 4800, Christchurch, New Zealand \\
% ...
$^{22}$ Niels Bohr Institute, University of Copenhagen, DK-2100 Copenhagen, Denmark \\
% ...
$^{28}$ D{\'e}partement de physique nucl{\'e}aire et corpusculaire, Universit{\'e} de Gen{\`e}ve, CH-1211 Gen{\`e}ve, Switzerland \\
% ...
$^{31}$ Karlsruhe Institute of Technology, Institute for Astroparticle Physics, D-76021 Karlsruhe, Germany  \\
% ...
$^{42}$ Bartol Research Institute and Dept. of Physics and Astronomy, University of Delaware, Newark, DE 19716, USA \\
% ...
$^{50}$ Oskar Klein Centre and Dept. of Physics, Stockholm University, SE-10691 Stockholm, Sweden \\
% ...
$^{59}$ DESY, D-15738 Zeuthen, Germany \\
% ...

\subsection*{Acknowledgements}

\noindent
USA {\textendash} U.S. National Science Foundation-Office of Polar Programs,
U.S. National Science Foundation-Physics Division,
U.S. National Science Foundation-EPSCoR,
Wisconsin Alumni Research Foundation,
Center for High Throughput Computing (CHTC) at the University of Wisconsin{\textendash}Madison,
Open Science Grid (OSG),
Extreme Science and Engineering Discovery Environment (XSEDE),
Frontera computing project at the Texas Advanced Computing Center,
U.S. Department of Energy-National Energy Research Scientific Computing Center,
Particle astrophysics research computing center at the University of Maryland,
Institute for Cyber-Enabled Research at Michigan State University,
and Astroparticle physics computational facility at Marquette University;
Belgium {\textendash} Funds for Scientific Research (FRS-FNRS and FWO),
FWO Odysseus and Big Science programmes,
and Belgian Federal Science Policy Office (Belspo);
Germany {\textendash} Bundesministerium f{\"u}r Bildung und Forschung (BMBF),
Deutsche Forschungsgemeinschaft (DFG),
Helmholtz Alliance for Astroparticle Physics (HAP),
Initiative and Networking Fund of the Helmholtz Association,
Deutsches Elektronen Synchrotron (DESY),
and High Performance Computing cluster of the RWTH Aachen;
Sweden {\textendash} Swedish Research Council,
Swedish Polar Research Secretariat,
Swedish National Infrastructure for Computing (SNIC),
and Knut and Alice Wallenberg Foundation;
Australia {\textendash} Australian Research Council;
Canada {\textendash} Natural Sciences and Engineering Research Council of Canada,
Calcul Qu{\'e}bec, Compute Ontario, Canada Foundation for Innovation, WestGrid, and Compute Canada;
Denmark {\textendash} Villum Fonden and Carlsberg Foundation;
New Zealand {\textendash} Marsden Fund;
Japan {\textendash} Japan Society for Promotion of Science (JSPS)
and Institute for Global Prominent Research (IGPR) of Chiba University;
Korea {\textendash} National Research Foundation of Korea (NRF);
Switzerland {\textendash} Swiss National Science Foundation (SNSF);
United Kingdom {\textendash} Department of Physics, University of Oxford.

\end{document}